\documentclass[runningheads]{llncs}
\usepackage[T1]{fontenc}
\usepackage{cite}
\usepackage{amsmath,amssymb,amsfonts}
\usepackage{algorithmic}
\usepackage{graphicx}
\usepackage{textcomp}
\usepackage{tikz}
\usepackage{tcolorbox}
\usepackage{todonotes}
\usepackage{adjustbox}
\usepackage{booktabs}
\usepackage{caption}
\usetikzlibrary{positioning}
\usetikzlibrary{tikzmark} 
\usetikzlibrary{fit}
\usetikzlibrary{matrix}
\usepackage[ruled,vlined]{algorithm2e}

\definecolor{y}{RGB}{255, 250, 230}
\begin{document}

\title{Keeping track of errors: A study of SHACL-DS for RDF dataset validation on the ERA RINF Knowledge Graph}

\author{
    Davan Chiem Dao\inst{1}\orcidID{0009-0004-8139-1927} \and
    Ghislain Atemezing\inst{2}\orcidID{0000-0003-1562-6922} \and
    Christophe Debruyne\inst{1}\orcidID{0000-0003-4734-3847}
}

\authorrunning{D. Chiem Dao et al.}

\institute{
    Montefiore Institute, University of Liège, Liège, Belgium\\
    \email{\{davan.chiemdao,c.debruyne\}@uliege.be}
\and
    European Union Agency for Railways (ERA), Valenciennes, France\\
    \email{ghislain.atemezing@era.europa.eu}
}

\maketitle

\begin{abstract}
SHACL-DS extends SHACL for RDF dataset validation by introducing declarative targeting of named graphs and graph combinations, but has not yet been demonstrated and assessed on a real, large-scale Knowledge Graph (KG). In this paper, we apply the SHACL-DS approach to validate its use on such a KG.
We apply SHACL-DS to the European Railway Infrastructure (ERA RINF) KG, a large-scale RDF dataset in which 56 infrastructure managers contribute data to dedicated named graphs.
We migrate the ERA-RINF shapes to SHACL-DS using two strategies and evaluate their performance using a TopBraid SHACL-DS implementation developed for this study. We compare the performance against the SHACL approach, which "flattens" all graphs into a single data graph.
Both strategies produce the same results and are faster than the SHACL baseline.
Not only do we demonstrate that SHACL-DS is at least as expressive as SHACL, but SHACL-DS also allows the validation scope to be declared inside the shapes artefact, enforces triple provenance through \texttt{GRAPH} clauses, enriches validation reports with per-graph annotations, and enables shape organisation across named shapes graphs.
\end{abstract}

\section{Introduction}
\label{sec:introduction}

The Shapes Constraint Language (SHACL)~\cite{shacl} has become the standard for validating RDF graphs, but it was designed for a flat, single-graph data model.
In practice, many knowledge graphs are federated datasets composed of multiple named graphs contributed by different parties, each responsible for a distinct portion of the data.
SHACL cannot natively express which graphs should participate in validation, let alone enforce that certain triples come from specific, authoritative graphs.
The usual workaround is to merge selected named graphs into a single target graph before validation, a step that must be handled by bespoke code external to the shapes graph, which obscures triple provenance.

SHACL-DS~\cite{shaclds} extends SHACL for RDF datasets by introducing declarative targeting of named graphs and combinations thereof, directly inside the shapes artefact.
While the specification defines the conceptual framework, its applicability to a large, real-world federated knowledge graph has not yet been evaluated.

The ERA RINF Knowledge Graph~\cite{toledo2025} is a large-scale dataset published by the European Union Agency for Railways (ERA) that comprises 56 infrastructure managers\footnote{\url{https://rinf.data.era.europa.eu/dataset-explorer}} each contributing data in a dedicated named graph, publicly available\footnote{\url{https://rinf.data.era.europa.eu/}}.
ERA currently validates this dataset with bespoke scripts that merge graphs into a single data graph before applying SHACL rules, resulting in a loss of provenance, e.g., determining which graphs introduced errors.
ERA therefore represents both a challenging and highly relevant test case for SHACL-DS.

In this paper, we migrate the ERA-RINF shapes to SHACL-DS and evaluate performance using the TopBraid SHACL-DS implementation to answer the following two research questions:

\begin{description}
  \item[RQ1 - Expressivity.] How does SHACL-DS compare to SHACL in terms of expressivity?
  \begin{description}
    \item[RQ1a - Equivalence.] Can SHACL-DS produce the same validation results as SHACL?
    \item[RQ1b - Added value.] What can SHACL-DS express beyond SHACL?
    \item[RQ1c - Performance.] What is the performance difference between SHACL and SHACL-DS?
  \end{description}
  \item[RQ2 - Practical Value.] What is the practical value of SHACL-DS for the ERA KG, and what effort does it require?
  \begin{description}
    \item[RQ2a - Added value.] What does SHACL-DS enable for the ERA KG that SHACL cannot?
    \item[RQ2b - Migration effort.] What is required to migrate existing ERA shapes to SHACL-DS, and how difficult is it?
  \end{description}
\end{description}

Section~\ref{sec:background} introduces SHACL-DS, the ERA RINF Knowledge Graph, and the existing ERA shapes.
Section~\ref{sec:related} discusses related work.
Section~\ref{sec:migration} describes the migration of ERA shapes to SHACL-DS.
Section~\ref{sec:reorganisation} presents the shape reorganisation enabled by SHACL-DS.
Section~\ref{sec:performance} reports performance results across all six configurations.
Section~\ref{sec:results} answers the two research questions.
Section~\ref{sec:discussion} discusses open challenges and directions for future work.
Section~\ref{sec:conclusion} concludes.

\section{Background and Related Work}
\label{sec:background}

\subsection{SHACL and SHACL-DS}
\label{sec:shacl-ds}

SHACL~\cite{shacl} is a W3C Recommendation for validating RDF graphs.
A \emph{shapes graph} contains a set of \emph{shapes}, each selecting a set of \emph{focus nodes} via target declarations and applying \emph{constraints} to them.
Violations are reported in a \emph{validation report}.
Constraints may be expressed as SHACL-core property constraints or as SPARQL-based constraints.

SHACL operates on a single RDF graph. When data is organised across multiple named graphs, practitioners must either "flatten" the dataset into a single graph, thereby losing provenance, or rely on bespoke preprocessing that embeds validation logic outside the shapes graph~\cite{shaclds}. A problem observed at the European Union Agency for Railways, which is introduced later in this paper.

SHACL-DS~\cite{shaclds} extends SHACL to validate RDF datasets directly, treating whole graphs (i.e., default graphs, named graphs, or combinations of graphs) as first-class validation objects.
It introduces four mechanisms:

\smallskip\noindent\textbf{Shapes Dataset.}
A \emph{Shapes Dataset} is an RDF dataset whose named graphs contain SHACL shapes.
The default graph holds targeting declarations that associate each shapes graph with one or more \emph{focus graphs} drawn or derived from the data dataset.

\smallskip\noindent\textbf{Target Graph Declarations} (\texttt{shds:targetGraph}).
A shapes graph is associated with data graphs by IRI.

\smallskip\noindent\textbf{Target Graph Combinations} (\texttt{shds:targetGraphCombination}).
A new focus graph is derived from existing graphs via the set operators \texttt{shds:and} (intersection), \texttt{shds:or} (union), and \texttt{shds:minus} (difference).
Each combination declaration yields exactly one combined focus graph.

\smallskip\noindent\textbf{Dataset Views}
For each (shapes graph, focus graph) pair, SHACL-DS constructs an \emph{evaluation dataset} in which the focus graph becomes the default graph and the original default graph is preserved in a named graph with IRI \texttt{shds:default}.
SHACL leaves the behaviour of \texttt{GRAPH}, \texttt{FROM}, and \texttt{FROM NAMED} in SPARQL-based constraints underspecified for RDF datasets; SHACL-DS resolves this by making these clauses operate on the full \emph{evaluation dataset}, thus allowing cross-graph constraints to have well-defined semantics.
\begin{sloppypar}
\smallskip\noindent\textbf{Validation report extensions}
Two new predicates extend SHACL's vocabulary for validation reports to provide provenance to resulting errors: \texttt{shds:focusGraph} identifies the data graph in which the focus node originated, and \texttt{shds:sourceShapeGraph} identifies the shapes graph that generated the error, the dataset-level analogues of \texttt{sh:focusNode} and \texttt{sh:sourceShape}.
\end{sloppypar}

\subsection{The ERA RINF Knowledge Graph}
\label{sec:era-kg}
\begin{sloppypar}
The European Union Agency for Railways (ERA) maintains the Register of Infrastructure (RINF), a system collecting railway infrastructure data from Infrastructure Managers (IMs) and National Registration Entities (NREs) across EU member states~\cite{rojas2021}.
Since 2021, ERA has published this data as an RDF Knowledge Graph, making it queryable via a SPARQL endpoint and reusable for applications such as Route Compatibility Checks~\cite{toledo2025}, with further usage such as routing calculation on the network with a micro level description. 
The ERA ontology, explicitly mentioned in the RINF regulation\footnote{\url{http://data.europa.eu/eli/reg\_impl/2023/1694/oj}}, formalizes the KG with approximately 76 classes  (e.g., \texttt{era:OperationalPoint}), 600 properties (Objet/Data properties), 52 annotation properties,  and more than 80 SKOS concept schemes for coded parameter values~\cite{toledo2025,erav322,erabenchmark}.
\end{sloppypar}

\begin{sloppypar}

The ERA KG is structured as an RDF dataset. We use the February 2026 dump (v2026-0217)~\cite{erakg}, which comprises 56 named graphs.
 
The majority are \emph{operator graphs}: named graphs each contributed independently by an infrastructure manager and identified by its company code.
Their sizes vary considerably, from fewer than 1,000 triples to over 10 million (France and Germany having the biggest network).
Beyond operator graphs, the dataset includes ERA-managed shared reference graphs: ontology graphs, SKOS controlled vocabulary graphs, a countries graph, a reference borders graph, SHACL validation graphs and a dataset metadata graph.
Some ERA-managed graphs exist in multiple versions: the ontology and SKOS vocabulary each appear in three named graphs across two versions, v3.0.1 duplicated as both \texttt{era-g:ontology} and \texttt{era-rinf:ontology}, and v3.1.5 as \texttt{era-315:ontology} (respectively \texttt{era-g:skos}, \texttt{era-rinf:skos}, and \texttt{era-315:skos}).
Table~\ref{tab:graphs} summarises the graph categories. Note that namespace prefix bindings used throughout this paper are listed in Appendix~\ref{sec:namespaces}.
\end{sloppypar}
\vspace{-2.7em}
\begin{table}[h]
\caption{Named graph categories in the ERA RINF dataset (56 graphs total)}
\label{tab:graphs}
\centering
\small
\begin{tabular}{|l|l|l|}
\hline
\textbf{Category} & \textbf{Example IRI} & \textbf{Count} \\
\hline
Operator (rinf) & \texttt{era-rinf:0085} & 44 \\
Ontology & \texttt{era-g:ontology}& 3 \\
SKOS vocabulary & \texttt{era-g:skos} & 3 \\
Countries & \texttt{era-g:countries} & 1 \\
Borders & \texttt{era-g:borders} & 1 \\
SHACL shapes & \texttt{era-315:shacl} & 3 \\
Dataset metadata & \texttt{era-rinf:dataset} & 1 \\
\hline
\end{tabular}
\end{table}
\vspace{-1.5em}
The dataset totals approximately 33.6 million triples across all 56 named graphs.
Given this scale, its federated multi-contributor structure, and ontology versioning, the ERA KG constitutes a representative and challenging test case for dataset-level validation.

\subsection{ERA-RINF Shapes}
\label{sec:era-shapes}
\begin{sloppypar}

Data correctness is critical in the railway domain, and ERA publishes SHACL shapes to declaratively validate RINF data~\cite{toledo2025,erav322,erabenchmark}.
In this work, we use version 3.2.0~\cite{erav32}, the latest available at the time of the experiment, comprising 458 shapes (75 NodeShapes and 383 PropertyShapes).
\end{sloppypar}
Here is a representative property constraint: 
\begin{tcolorbox}[
  boxrule=0.5pt,
  colback=y,
  colframe=y,
  sharp corners,
  left=1pt, right=1pt, top=1pt, bottom=1pt
]
\scriptsize
\begin{verbatim}
era-sh:ContactLineSystemShape sh:property era-sh:MaximumTrainCurrent .
era-sh:MaximumTrainCurrenta sh:PropertyShape ;
 rdfs:comment "Indication of the maximum allowable train current"@en;
 sh:path era:maxTrainCurrent ;
 sh:datatype xsd:integer ;
 sh:pattern "^([1-9]\\d{0,3}|0)$" ;  
 sh:maxCount 1 ;
 sh:severity sh:Violation ;
 sh:message "maxTrainCurrent (1.1.1.2.2.2): Defines ...(truncated)"@en .
\end{verbatim}
\end{tcolorbox}

\texttt{era-sh:ContactLineSystemShape} attaches \texttt{era-sh:MaximumTrainCurrent}, which enforces that \texttt{era:maxTrainCurrent} holds at most one integer value matching a specific numeric pattern.

\begin{sloppypar}
Beyond SHACL-core constraints, the ERA shapes also use SHACL-SPARQL to express rules that cannot be captured with core constraint components alone.
Several such shapes validate taxonomy membership: they check that a property value belongs to the correct SKOS concept scheme by looking up \texttt{era:inSkosConceptScheme} in the ontology graph.
For example:
\end{sloppypar}

\begin{tcolorbox}[
  boxrule=0.5pt,
  colback=y,
  colframe=y,
  sharp corners,
  left=1pt, right=1pt, top=1pt, bottom=1pt
]

\scriptsize
\begin{verbatim}
era-sh:ETCSShape sh:sparql era-sh:EtcsMVersionSKOS.
era-sh:EtcsMVersionSKOS	a sh:SPARQLConstraint ;
 rdfs:comment "ETCS_M version according to SRS 7.5.1.9 "@en ;
 sh:message "Indication of the etcsMVersion(truncated)"@en ;
 sh:prefixes era:;
 sh:select """
 PREFIX era: <http://data.europa.eu/949/>
 PREFIX skos: <http://www.w3.org/2004/02/skos/core#>
 PREFIX rdfs: <http://www.w3.org/2000/01/rdf-schema#>
 SELECT $this  ?concept (era:etcsMVersion AS ?path)
 WHERE {
  $this era:etcsMVersion ?concept .
  era:etcsMVersion era:inSkosConceptScheme ?conceptScheme .
  FILTER NOT EXISTS{ ?concept skos:inScheme ?conceptScheme .} } """.
\end{verbatim}
\end{tcolorbox}

\texttt{era-sh:EtcsMVersionSKOS} retrieves the concept scheme from the ontology (line~13), and  checks SKOS membership(line~14); both rely on ontology and SKOS triples being present in the same graph as the RINF data.

\subsection{ERA-SHACL-Benchmark}
\label{sec:related}

The ERA-SHACL-Benchmark~\cite{erabenchmark} evaluates SHACL validator performance on the ERA KG by merging all graphs into a single data graph. While it uses an older dataset dump and an older version of the ERA shapes, the order of magnitude and the nature of the shapes are comparable to those used in this work, making it a useful performance reference. It does not address dataset-level validation.

\section{From SHACL to SHACL-DS}
\label{sec:migration}

Validating the ERA KG with SHACL requires deciding which graphs to include and how to combine them.
Two prior works provide partial answers, but neither fully specifies the validation scope.
The ERA-SHACL-Benchmark~\cite{erabenchmark} builds its validation input via a shell script that strips named graph identifiers and merges operator data with vocabulary files, discarding all graph structure; while the benchmark does not claim this merged model to be the correct validation scope, one could assume it reflects the intended approach.
The RINF System~\cite{toledo2025} describes a pipeline in which SHACL validation is performed before the KG is published, but does not specify whether each operator graph is validated in isolation or combined with reference graphs.

In both cases, the validation scope is encoded outside the shapes, hidden in bespoke code, which is a limitation SHACL-DS addresses~\cite{shaclds}.

The reality likely sits between the two approaches, but as neither source fully specifies it. We define the SHACL baseline as the union of the 44 operator graphs together with \texttt{era-g:ontology}, \texttt{era-g:skos}, \texttt{era-g:countries}, and \texttt{era-g:borders}, which we consider to be the sole graphs required to validate railway infrastructure data, and use it as the reference point for all comparisons. We refer to the union of all 56 named graphs as \emph{SHACL-full} when needed.
In the following sections, we show that migrating to SHACL-DS offers different ways to specify the validation scope declaratively, each with distinct trade-offs.
In all configurations, we use the ERA SHACL shapes version 3.2.0~\cite{erav32} as the shapes graph \texttt{era-sh:sg-rinf}, and the ERA KG dataset dump \cite{erakg} is kept as-is with no modification or preprocessing.

\subsection{Combination Strategy}
\label{sec:combo}

Since SHACL-DS is a backward-compatible extension of SHACL~\cite{shaclds}, any data graph that would be produced by some preprocessing and then validated with SHACL can equally be validated with SHACL-DS with the same preprocessing, directly answering RQ1.
What SHACL-DS adds is the ability to express declaratively the target of a shapes graph as a combination of named graphs, replacing the graph selection logic previously hidden in external scripts.
In particular, the SHACL-full configuration can be expressed declaratively as:

\begin{tcolorbox}[
  boxrule=0.5pt,
  colback=y,
  colframe=y,
  sharp corners,
  left=1pt, right=1pt, top=1pt, bottom=1pt
]
\scriptsize
\begin{verbatim}
era-sh:sg-rinf shds:targetGraphCombination [ shds:or ( shds:all ) ] .
\end{verbatim}
\end{tcolorbox}

This would produce the same validation report as SHACL-full, apart from SHACL-DS-specific report annotations.\footnote{The same could be done for the SHACL baseline by explicitly enumerating the 44 operator graphs and the four reference graphs in a single combination. }

The Combination Strategy goes further by targeting each operator graph individually.
For each of the 44 operator graphs, a target graph combination is declared that unions the operator graph with the ERA-managed reference graphs (ontology, SKOS, countries, and borders).
The shared reference graphs are factored out as a reusable blank node:

\begin{tcolorbox}[
  boxrule=0.5pt,
  colback=y,
  colframe=y,
  sharp corners,
  left=1pt, right=1pt, top=1pt, bottom=1pt
]

\scriptsize
\begin{verbatim}_:refGraphs shds:or ( era-g:ontology era-g:skos era-g:countries era-g:borders ) .
era-sh:sg-rinf shds:targetGraphCombination [ shds:or ( era-rinf:0080 _:refGraphs ) ] .
era-sh:sg-rinf shds:targetGraphCombination [ shds:or ( era-rinf:0085 _:refGraphs ) ] .
# one declaration per operator graph
\end{verbatim}
\end{tcolorbox}

The required graphs for each operator are now specified declaratively in the shapes dataset, making the previously undocumented validation scope explicit.

\subsection{Target Strategy}
\label{sec:target-graph}

Where the Combination Strategy merges reference graphs into the focus graph, the Target Strategy associates \texttt{era-sh:sg-rinf} with each operator graph directly.
Each operator graph serves as the focus graph, containing only that operator's data and constraints that require triples from other graphs must be expressed as SPARQL-based constraints with explicit \texttt{GRAPH} clauses.

To achieve this, a single \texttt{shds:targetGraphPattern} declaration matching the 4-character alphanumeric company code suffices rather than enumerating all 44 operator graphs individually:

\begin{tcolorbox}[
  boxrule=0.5pt,
  colback=y,
  colframe=y,
  sharp corners,
  left=1pt, right=1pt, top=1pt, bottom=1pt
]
\scriptsize
\begin{verbatim}
era-sh:sg-rinf shds:targetGraphPattern ".*/graph/rinf/[A-Z0-9]{4}$" .
\end{verbatim}
\end{tcolorbox}

For each matching graph, SHACL-DS constructs an evaluation dataset in which that graph is the default graph and all other named graphs remain accessible by their IRIs~\cite{shaclds}, making cross-graph \texttt{GRAPH} lookups well-defined.\footnote{The Combination Strategy also uses evaluation datasets, but since the combined focus graph already merges all required triples, the original shapes require no \texttt{GRAPH} clauses and the full dataset view is not needed.}

\begin{sloppypar}
    
Migration therefore requires identifying shapes that depend on data from other graphs and rewriting them as SPARQL-based constraints with explicit \texttt{GRAPH} clauses.
Many of the ERA SPARQL constraints require data from the ontology, SKOS, or countries graphs and must therefore be rewritten.
The \texttt{era-sh:EtcsMVersionSKOS} shape from Section \ref{sec:era-shapes} is a representative example: it retrieves the concept scheme from the ontology (line~13) and checks SKOS membership (line~14) using bare triple patterns that only work when those triples are in the same graph as the focus node.
In this strategy, both lookups are wrapped in explicit \texttt{GRAPH} clauses and the SPARQL query is rewritten as:
\end{sloppypar}

\begin{tcolorbox}[
  boxrule=0.5pt,
  colback=y,
  colframe=y,
  sharp corners,
  left=1pt, right=1pt, top=1pt, bottom=1pt
]
\scriptsize
\begin{verbatim}
SELECT $this ?concept (era:etcsMVersion AS ?path) WHERE {
  $this era:etcsMVersion ?concept .
  GRAPH era-g:ontology { era:etcsMVersion era:inSkosConceptScheme ?conceptScheme . }
  FILTER NOT EXISTS { GRAPH era-g:skos { ?concept skos:inScheme ?conceptScheme . } } }
\end{verbatim}
\end{tcolorbox}

A subtler case arises with SHACL-core constraints that implicitly depend on ontology triples.
\texttt{era-sh:NotApplicableProperty} checks that each value of \texttt{era:notApplicable} is a data property or an object property, using a SHACL-core class constraints:

\begin{tcolorbox}[
  boxrule=0.5pt,
  colback=y,
  colframe=y,
  sharp corners,
  left=1pt, right=1pt, top=1pt, bottom=1pt
]
\scriptsize
\begin{verbatim}
sh:or ( [ sh:class owl:ObjectProperty ] [ sh:class owl:DatatypeProperty ] ) .
\end{verbatim}
\end{tcolorbox}

Since those class assertions live in the ontology graph, this constraint incorrectly fails for all values when the ontology is absent from the focus graph.
The constraint must therefore be rewritten as a SPARQL constraint with explicit \texttt{GRAPH} clauses:

\begin{tcolorbox}[
  boxrule=0.5pt,
  colback=y,
  colframe=y,
  sharp corners,
  left=1pt, right=1pt, top=1pt, bottom=1pt
]
\scriptsize
\begin{verbatim}
sh:sparql [ sh:select """
  SELECT DISTINCT $this WHERE {
    $this era:notApplicable ?property .
    FILTER NOT EXISTS {
      { GRAPH era-ng:ontology { ?property a owl:ObjectProperty . } }
      UNION
      { GRAPH era-ng:ontology { ?property a owl:DatatypeProperty . } } }  } """ ] .
\end{verbatim}
\end{tcolorbox}

\begin{sloppypar}

These two examples illustrate the main migration effort in the Target Strategy: identifying shapes with implicit cross-graph dependencies and rewriting them as SPARQL-based constraints.
Most SHACL-core shapes that do not depend on other graphs migrate without modification, requiring only the \texttt{shds:targetGraphPattern} declaration.
\end{sloppypar}

\subsection{Migration Validation}
\label{sec:migration-validation}

To verify that the migration produces correct results, we compare the two SHACL-DS strategies against each other and against the SHACL baseline.
Comparison is done by grouping errors by \texttt{sh:sourceShape} and comparing counts.
While it does not guarantee that each individual error is identical across approaches (equal counts for a shape could mask differences in which focus nodes are flagged, for instance, if gains and losses for the same shape cancel out; or, if a shape that would require rewriting is never triggered because the relevant data is absent from the graph), it is a good proxy for equivalence at the scale of 1.2 million errors.

\paragraph{Combination Strategy vs.\ Target Strategy.}
Comparing error counts per shape between the two strategies was used throughout the Target Strategy migration. Any shape reporting different counts indicated a cross-graph dependency that had not yet been made explicit and was rewritten using \texttt{GRAPH} clauses until both strategies agreed.
After migration, both strategies report identical counts across all shapes, confirming they achieve the same validation.

\paragraph{SHACL-DS vs.\ SHACL baseline.}
Having established that both strategies agree, we now compare them against the SHACL baseline.
Table~\ref{tab:counts} reports error counts for both approaches.

\begin{table}[h]
\caption{Error counts: SHACL vs.\ SHACL-DS}
\label{tab:counts}
\centering
\small
\begin{tabular}{|l|r|r|}
\hline
 & \textbf{SHACL} & \textbf{SHACL-DS} \\
\hline
Total errors & 1,217,993 & 1,250,354 \\
Total duplicated errors removed & 378 & 32,712 \\
Total duplicated errors    & 1,217,615 & 1,217,642 \\
\hline
Shapes with identical count & \multicolumn{2}{c|}{51} \\
Shapes differing          & \multicolumn{2}{c|}{3} \\
\hline
\end{tabular}
\end{table}

The raw counts differ by 32,361, mainly due to per-operator repetition: both strategies validate each operator graph independently, so an error found in data shared across operators is reported once per operator graph, whereas in the SHACL baseline, the same triples are merged and appear only once.
We therefore apply deduplication, which collapses errors that share the same SHACL validation result annotations to compare counts.\footnote{The 378 errors removed from the SHACL report are a separate artefact: a SPARQL constraint missing a \texttt{DISTINCT} keyword causes some results to appear twice.}
While the per-operator duplicates some reported errors, each carries the \texttt{shds:focusGraph} annotation identifying which operator graph produced it, allowing ERA to directly notify the responsible infrastructure manager without any post-processing.
With SHACL, identifying the responsible operator from a merged-model report would require a separate SPARQL query against the original dataset.

After deduplication, the residual gap is 27 errors across 3 shapes (1,200 more in SHACL-DS, 1,070 and 103 more in SHACL, netting to 27), making the two approaches nearly equivalent by error count.
These differences could, in principle, indicate a migration error, but inspection reveals they all stem from shapes with cross-graph interactions between operator graphs, where errors can appear or disappear depending on whether validation is per-operator (SHACL-DS) or merged (SHACL).
For example, two Swiss operator graphs overlap on the same sections of line: one declares a reference that the other lacks.
In the SHACL-DS approach, the operator missing the reference receives an error. In the merged model, the reference is borrowed from the other graph, and the constraint passes.
The inverse is also observed: sections where both operators declare conflicting values for the same property pass in per-operator validation (each graph contains one value) but fail in the merged model where both values are visible simultaneously.
As noted earlier, neither prior work fully specifies the correct validation scope for the ERA KG, so it is not possible to determine which behaviour is correct.
SHACL-DS is capable of replicating either; the choice of merging or not is a modelling decision, not a limitation of the framework.
We discuss the implications further in Section~\ref{sec:discussion}.

\section{Shape Reorganisation}
\label{sec:reorganisation}

SHACL operates on a single flat shapes graph, regardless of which data the individual shapes target.
Just as selecting which data graphs to include requires bespoke preprocessing code in SHACL, managing multiple shapes graphs also requires external orchestration to decide which shapes to apply to which data.
The SHACL-DS Shapes Dataset makes both explicit: named shapes graphs enable shapes to be organised by purpose, each with its own declarative target declarations.
During migration, it became apparent that the ERA shapes contained constraints that were not about RINF infrastructure data at all, but about the structure of the ontology itself.
This was not accidental: a comment present in the ERA shapes file read:

\vspace{2mm}
\texttt{\# ERA-ONTO part, to be extracted + smart merge in the future}\footnote{\url{https://gitlab.com/era-europa-eu/public/interoperable-data-programme/era-ontology/era-ontology/-/blob/712b9475fe8a4d194457fe54e82efe1db2e72483/era-shacl/ERA-RINF-shapes.ttl\#L12392}}
\vspace{2mm}

This shows explicitly acknowledgment that ontology-specific shapes should be separated but that the tooling to do so did not exist.

\begin{sloppypar}
SHACL-DS directly addresses this need.
Three additional named shapes graphs were created, each targeting a specific category of graphs via \texttt{shds:targetGraphPattern}:
\end{sloppypar}

\begin{tcolorbox}[
  boxrule=0.5pt,
  colback=y,
  colframe=y,
  sharp corners,
  left=1pt, right=1pt, top=1pt, bottom=1pt
]
\scriptsize
\begin{verbatim}
era-sh:sg-ont   shds:targetGraphPattern ".*/ontology$" .
era-sh:sg-skos  shds:targetGraphPattern ".*/skos$" .
era-sh:sg-shacl shds:targetGraphPattern ".*/shacl$" .
\end{verbatim}
\end{tcolorbox}
\
\begin{sloppypar}
    
\texttt{era-sh:sg-ont} targets all ontology graphs and contains shapes extracted from the ERA SHACL shapes graph that validate ontology structure.
\texttt{era-sh:sg-skos} targets all SKOS vocabulary graphs and contains shapes that validate concept scheme structure.\footnote{These shapes were retrieved from an older version of the ERA shapes repository for illustration purposes, as they appear to have been removed from the current version.}
\texttt{era-sh:sg-shacl} targets all SHACL shapes graphs, enabling SHACL-on-SHACL validation.
\end{sloppypar}

This reorganisation illustrates how SHACL-DS enables shapes to be organised and their application scope to be declared explicitly, rather than left implicit in external scripts or mixed together in a single flat graph.

\section{Performance Evaluation}
\label{sec:performance}

We evaluate whether SHACL-DS is viable at the scale of the ERA KG by comparing validation times across several configurations.
The initial SHACL-DS prototype relied on and extended dotNetRDF~\cite{shaclds,dotnetrdf_shacl}, which the ERA-SHACL-Benchmark~\cite{erabenchmark} ranks among the slower SHACL implementations, and proved too slow and memory-intensive on the full ERA dataset.
We therefore implemented SHACL-DS as an extension of the TopBraid SHACL API~\cite{topbraid_shacl}, which ranks among the fastest implementations in the benchmark.
All benchmarks reported here were run on this TopBraid-based implementation.

\begin{figure}[h]
\centering
\includegraphics[width=\linewidth]{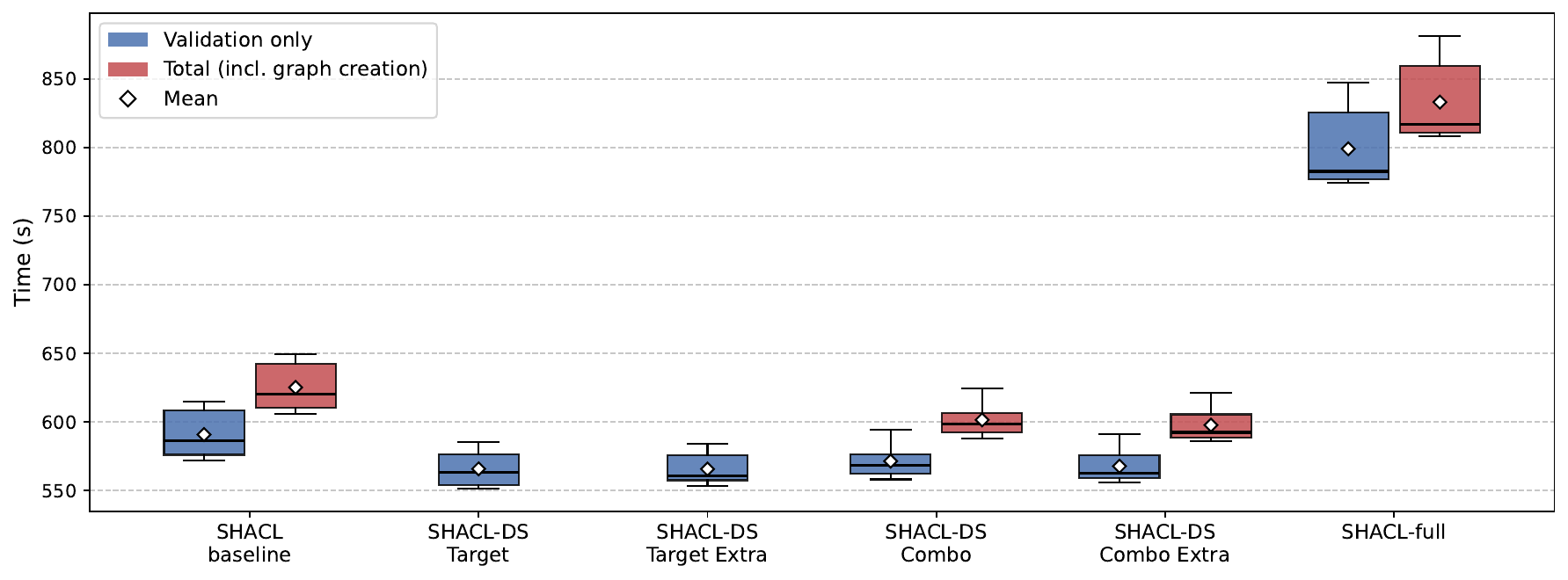}
\caption{Validation time distribution per configuration over 10 runs.}
\label{fig:perf}
\end{figure}

All configurations run against the same in-memory dataset: the ERA KG is loaded from disk once (91\,s) before any timing begins, and this cost is excluded from all measurements.
Each configuration loads its shapes file fresh per run and measures graph creation time (merge or view creation) and validation separately.
Each configuration is run 10 times; Figure~\ref{fig:perf} shows the distribution of validation times, and Table~\ref{tab:violations} reports the error counts. The full numerical results and machine configuration are provided in Appendix~\ref{sec:results-table} and \ref{sec:machine}.

All configurations reach a peak JVM heap usage of about 32\,GB, similar to the benchmark results, dominated by the in-memory dataset. 

\begin{table}[h]
\caption{Reported errors per configuration}
\label{tab:violations}
\centering
\small
\begin{tabular}{|l|r|}
\hline
\textbf{Configuration} & \textbf{Errors} \\
\hline

SHACL baseline                 & 1,217,993 \\
SHACL-DS Target/Combination Strategy       & 1,250,354 \\
SHACL-DS Target/Combination Extra Strategy  & 1,255,571 \\
SHACL-full                     & 2,770,068 \\
\hline
\end{tabular}
\end{table}

One might expect SHACL-DS to be slower than SHACL given that it operates on a more complex structure (RDF dataset vs RDF graph). However, the results show the opposite.
The fair comparison is between the SHACL baseline, the Target Strategy, and the Combination Strategy, as all three validate the same amount of data: the 44 operator graphs combined with the four reference graphs.
The Extra configurations extend this with additional shapes graphs described in Section \ref{sec:reorganisation}, targeting the ontology, SKOS, and SHACL graphs.

Both SHACL-DS strategies complete on average end-to-end validation faster than the SHACL baseline.
The Target Strategy is the fastest overall. This is mostly because it does not require creating any intermediate graph.
Looking at validation time alone, the Target Strategy (566\,s) and the Combination Strategy (568\,s) are both faster than the SHACL baseline (591\,s). 
While SHACL-DS must perform multiple SHACL validations (one per operator graph), each operates on a smaller search space, which we believe more than compensates for the iteration overhead.
For the Target Strategy specifically, constraints without \texttt{GRAPH} clauses see only one operator graph. Those with \texttt{GRAPH} clauses do access the full evaluation dataset\footnote{including the 7 graphs excluded from the SHACL baseline}, but we believe the low selectivity of these clauses keeps the overall cost low.
The gain could in principle be even larger, SHACL-DS reports more errors due to per-operator repetition, and the current TopBraid implementation copies each per-graph result into a merged report after every targeted graph validation, which may introduce an additional cost. 

The Combination Strategy requires graphs combinations creation steps (30\,s), comparable in cost to the graph merge required by the SHACL baseline (34\,s), and achieves a slightly faster validation time.

The Extra configurations add modest overhead over their base counterparts, the additional shapes graphs target contain about ~50k triples (the ontology, SKOS, and SHACL graphs) compared to the ~33 million triples already validated, so the extra cost is negligible.

SHACL-full is by far the slowest and reports more than twice the amount of the baseline despite differing by about only ~50k triples. The additional graphs introduce multiple ontology and SKOS versions with potential contradictions, illustrating the need for explicitly declaring the validation scope.

\section{Results}
\label{sec:results}

\subsection{RQ1 - Expressivity}

\begin{sloppypar}
\smallskip\noindent\textit{RQ1a - Equivalence.}
SHACL-DS, as a backward-compatible extension of SHACL, can validate any graph that SHACL can (Section~\ref{sec:combo}).
Beyond backward compatibility, SHACL-DS can also declaratively express validation that requires a dataset to be "flattened" into a single graph with a target graph combination that is the union of all graphs.
\end{sloppypar}
We also show that two distinct SHACL-DS strategies, the Combination and Target Strategies, each can produce the same error report as the SHACL baseline\footnote{Solutions to the residual difference of 27 errors is discussed further in Section~\ref{sec:discussion}} (Section \ref{sec:migration-validation}).
Note that depending on the use case, the two strategies could be combined within the same shapes dataset: per-operator isolation for most shapes and a combination for those that require cross-operator visibility.

\begin{sloppypar}
\smallskip\noindent\textit{RQ1b - Added value.}
Beyond replicating SHACL, SHACL-DS enables several capabilities that SHACL cannot express at all.SHACL-DS allows the validation scope to be declared explicitly inside the shapes dataset, rather than encoded in external scripts (Sections~\ref{sec:combo}~and~\ref{sec:target-graph}).
In this work, we only used \texttt{shds:or} to build focus graphs, but SHACL-DS also provides \texttt{shds:and} (intersection) and \texttt{shds:minus} (difference), enabling any validation scope that can be expressed as a combination of named graphs through set operations.
\end{sloppypar}

The Target Strategy, through explicit \texttt{GRAPH} clauses, allows constraints to assert that triples come from a specific named graph.
When graphs are merged, whether through the Combination Strategy or as preprocessing of SHACL, graph provenance is lost in the focus graph: a triple required to satisfy constraint could originate from any of the merged graphs, with no way to enforce which one it should come from.

SHACL-DS also extends SHACL's validation report with two additional annotations: \texttt{shds:focusGraph} identifies the data graph in which the focus node originated, and \texttt{shds:sourceShapeGraph} identifies the shapes graph that produced the error.
While this means errors shared across focus graphs are reported multiple times (Section~\ref{sec:migration-validation}), each copy carries a \texttt{shds:focusGraph} annotation identifying exactly which source graph produced it.

\begin{sloppypar}
Finally, SHACL-DS allows shapes to be organised across named shapes graphs, each targeting a specific category of data, leveraging the knowledge organisation already present in the dataset (Section~\ref{sec:reorganisation}).
\end{sloppypar}

\begin{sloppypar}
\smallskip\noindent\textit{RQ1c - Performance.}
Contrary to expectations, SHACL-DS is faster than SHACL on the ERA KG.
End-to-end, the Target Strategy and the Combination Strategy both outperform the SHACL baseline (Section \ref{sec:performance}).
\end{sloppypar}

We hypothesise that the speedup comes from iterating validations on a smaller search space.
The Target Strategy has additional advantages. Unlike the Combination Strategy and the SHACL baseline, it requires no intermediate graph creation before validation begins. We believe it also benefits from the low selectivity of \texttt{GRAPH}-based constraints, which target specific well-known graphs and therefore speed up those lookups rather than adding overhead.
However, these hypotheses have not been verified with targeted experiments, and further investigation is needed to confirm the actual sources of the speedup.

\subsection{RQ2 - Practical Value}

\smallskip\noindent\textit{RQ2a - Added value.}
SHACL-DS addresses several concrete issues in the ERA KG validation setup.

\textbf{Explicit validation scope.}
Prior ERA documentation did not formally specify which graphs should be included in validation, nor how they should be combined.
SHACL-DS target declarations would allow this to be made explicit.

\textbf{Triple provenance enforcement.}
A pain point in the ERA KG is that infrastructure managers sometimes unintentionally enrich the dataset by declaring resources using the \texttt{era:} namespace within their own operator graphs, for instance using a mistyped property name or creating their own \texttt{owl:Class} or \texttt{skos:Concept}.
In a merged model, such declarations become indistinguishable from ERA-managed ontology or SKOS data, allowing constraints that look up concept schemes or class memberships to pass spuriously against operator-supplied triples rather than the authoritative ERA reference graphs.
This is similar to the SHACL bypass issue identified in~\cite{shacl-bypass}, where losing provenance by merging graphs allows data to satisfy constraints it was not intended to satisfy.
The Target Strategy addresses this directly: by rewriting cross-graph lookups with explicit \texttt{GRAPH} clauses pinned to ERA-managed graphs, constraints can only be satisfied by triples from the intended source, regardless of what operators have declared in their own graphs.

\textbf{Per-operator error attribution.}
In the ERA KG, some triples, such as reference border points, appear in multiple operator graphs.
Once merged, these duplicates are indistinguishable from single-origin triples and the validation report cannot indicate which operator graph the violating triple came from.
SHACL-DS per-operator validation preserves this provenance, as each error carries a \texttt{shds:focusGraph} annotation identifying its source graph, allowing ERA to directly notify the responsible infrastructure manager without any post-processing.

\textbf{Shape organisation.}
The ERA shapes file contained a comment acknowledging that ontology-specific shapes should be extracted but that the tooling did not exist.
The SHACL-DS shapes dataset, through its named shapes graphs and declarative targeting, would allow this separation to be made explicit: ontology, SKOS, and SHACL-meta shapes could each be placed in a dedicated shapes graph with its own target declarations, addressing the issue.

\smallskip\noindent\textit{RQ2b: Migration effort.}
The migration effort depends heavily on the chosen strategy.
The Combination Strategy requires minimal changes to the shapes themselves: for each operator graph, one \texttt{shds:targetGraphCombination} declaration is added to the shapes dataset. Since the combined focus graph already contains all required triples, no shape rewriting is needed.
The Target Strategy is more involved. Every shape must be analysed to identify those that rely on triples from other graphs, which can be tedious for large shapes graphs.
In practice, we proceeded in two steps. Shapes following a recurring pattern, such as paired \texttt{GRAPH era-ng:ontology} and \texttt{GRAPH era-ng:skos} lookups (Section~\ref{sec:target-graph}), were rewritten with the help of an LLM. With an LLM, we applied the transformation to the pattern. We deemed this approach justifiable for the purpose of this study. 
Remaining shapes requiring changes were then discovered by comparing per-shape error counts against the Combination Strategy: any discrepancy indicated an unresolved cross-graph dependency.
A limitation of this approach is that shapes with cross-graph dependencies that currently produce no errors would not surface as discrepancies and could therefore be missed. 

\section{Discussion and Future Work}
\label{sec:discussion}

\smallskip\noindent\textit{Cross-operator constraints.}
As shown in Section~\ref{sec:migration-validation}, a small number of residual differences remain after deduplication, caused by shapes that produce different results depending on whether operator graphs are validated in isolation or merged.
If the correct behaviour requires seeing data from multiple operator graphs simultaneously, some solutions could be: merging all graphs into a single focus graph is the same as SHACL behaviour; extracting only the affected shapes into a dedicated shapes graph targeting the union of all graphs; or rewriting those constraints can be rewritten with \texttt{GRAPH} clauses that explicitly reference the required graphs.
Each option would affect performance differently, and the trade-offs would need to be evaluated in future work.

\smallskip\noindent\textit{SHACL ontology treatment.}
During the Target Strategy migration, one shape had to be rewritten as a SPARQL constraint with explicit \texttt{GRAPH} clauses in order to fetch class membership triples from the ontology for subclass checking.
The SHACL spec defines \texttt{sh:class} to walk \texttt{rdfs:subClassOf} chains, but only over triples already present in the data graph. If the ontology is in a separate named graph, those triples are not visible, and the constraint fails.
A contradiction arises for the ERA case: one would want combine the ontology and operator graph for class-walking, while also having the ERA-managed ontology separated from the operator graph and ensuring that some triples originate from it thanks to a \texttt{GRAPH} clause.
pySHACL \cite{pyshacl} allows loading an ontology separately via a dedicated parameter, making ontology triples available for inferencing without merging them into the focus graph.
A similar mechanism could be defined in SHACL-DS, though further investigation would be needed to assess its impact. 

\smallskip\noindent\textit{Performance.}
The performance results are specific to the TopBraid-based implementation and may not generalise to other future SHACL-DS processors. 
The observed speedup could depend on many implementation-specific factors such as dataset indexing, query planning for SPARQL-based constraints, or the cost of merging per-graph results, and may not transfer to processors with different architectures. 
We should experiment further to validate these hypotheses. 

\smallskip\noindent\textit{Federated validation.}
An extension of the provenance enforcement discussed above is the use of \texttt{SERVICE} clauses in SHACL-SPARQL constraints to query authoritative external endpoints directly. 
Rather than pinning \texttt{GRAPH} lookups to a local named graph that may become stale, a constraint could query a canonical endpoint that always reflects the current reference data, ensuring a single source of truth. 
How \texttt{SERVICE} behaves within SHACL-DS datasets is currently undefined and would require explicit specification.

\section{Conclusion}
\label{sec:conclusion}

We presented an evaluation of SHACL-DS on the ERA RINF Knowledge Graph, a large-scale dataset comprising 44 named graphs, each contributed by an infrastructure manager.
Starting from the existing ERA shapes, we designed and implemented two SHACL-DS migration strategies, validated their correctness, and measured performance across six configurations.

On expressivity (RQ1), we showed that SHACL-DS can fully replicate SHACL validation (RQ1a): both the Combination Strategy and the Target Strategy can produce identical error counts to the SHACL baseline.
Beyond equivalence, SHACL-DS allows the validation scope to be declared inside the shapes artefact, enables provenance-aware constraints through explicit \texttt{GRAPH} clauses, enriches the validation report with focus graph and source shapes graph annotations, and supports flexible shape organisation across named shapes graphs (RQ1b).
Contrary to expectations, SHACL-DS is faster than the SHACL baseline which merges all graphs on the ERA KG(RQ1c).

On practical value (RQ2), SHACL-DS addresses concrete limitations of the current ERA validation setup: it makes the validation scope explicit, enforces triple provenance to prevent data pollution by infrastructure managers from bypassing constraints, and preserves per-operator error attribution(RQ2a).
Migration difficulty varies by strategy: the Combination Strategy requires only target declarations and no shape rewriting, while the Target Strategy requires shape-level analysis and rewriting of cross-graph lookups which can be tedious even with LLM assistance (RQ2b).

Open challenges include the treatment of cross-operator constraints, the interaction between \texttt{sh:class} graph-walking and ontologies stored in separate named graphs, and the formal treatment of federated validation through \texttt{SERVICE} clauses.
These questions, together with the generalisation of the performance results beyond the TopBraid implementation, constitute directions for future work.

\section*{Acknowledgment}
We would like to thank Ashley Caselli, the main developer of the TopBraid SHACL API, for his encouragement and interest in incorporating SHACL-DS into the library.

\newpage
\paragraph*{Supplemental Material Statement:}
    
The SHACL-DS implementation, migrated ERA RINF shapes and benchmark runner are available at the following GitHub repository:
\begin{itemize}
\item[] \url{https://anonymous.4open.science/r/SHACL-DS-ERA-use-case-AC3F/}
\end{itemize}

\paragraph{Use of Generative AI:}
We used Claude (Anthropic) and Grammarly to assist with writing, proofreading, LaTeX formatting and condensing sections to meet the page limit. Claude was also used during the SHACL-DS TopBraid implmenetation and to support the migration of SHACL shapes to the SHACL-DS format. All scientific content, results, and conclusions were developed solely by the authors.


\appendix
\section{Namespaces}
\label{sec:namespaces}

Throughout the paper we use the following namespace prefix bindings: 

{\small
\begin{itemize}
\item \texttt{sh:}  \texttt{http://www.w3.org/ns/shacl\#}
\item \texttt{shds:} \texttt{http://www.w3id.org/shacl-ds\#}
\item \texttt{era-g:} \texttt{http://data.europa.eu/949/graph/}
\item \texttt{era-rinf:} \texttt{http://data.europa.eu/949/graph/rinf/}
\item \texttt{era-315:} \texttt{http://data.europa.eu/949/graph/v3-1-5/}
\item \texttt{era-sh:} \texttt{http://data.europa.eu/949/shapes/}
\item \texttt{rdfs:} \texttt{http://www.w3.org/2000/01/rdf-schema\#}
\item \texttt{xsd:} \texttt{http://www.w3.org/2001/XMLSchema}

\end{itemize}
}

\section{Benchmark results}
\label{sec:results-table}
\begin{table}[h]
\vspace{-3em}
\caption{Six-number summary (min, Q1, median, Q3, max, mean) of validation times over 10 runs per configuration on the ERA RINF dataset.}
\centering
\small
\begin{tabular}{l c c c c c c}
\toprule
\textbf{Configuration} & \textbf{Min (s)} & \textbf{Q1 (s)} & \textbf{Median (s)} & \textbf{Q3 (s)} & \textbf{Max (s)} & \textbf{Mean (s)} \\
\midrule
SHACL baseline & 571.8 & 575.9 & 586.1 & 610.8 & 615.0 & 590.9 \\
\quad + Merge & 606.1 & 610.2 & 620.4 & 645.1 & 649.2 & 625.2 \\
SHACL-DS Target & 551.4 & 553.4 & 563.1 & 576.9 & 585.4 & 565.8 \\
SHACL-DS Target Extra & 553.6 & 557.5 & 560.6 & 576.2 & 584.1 & 565.8 \\
SHACL-DS Combo & 558.2 & 562.1 & 568.7 & 577.4 & 594.3 & 571.5 \\
\quad + View creation & 588.2 & 592.1 & 598.7 & 607.4 & 624.3 & 601.5 \\
SHACL-DS Combo Extra & 556.2 & 557.9 & 562.6 & 576.2 & 591.5 & 567.8 \\
\quad + View creation & 586.1 & 587.8 & 592.5 & 606.1 & 621.4 & 597.8 \\
SHACL-full & 774.2 & 776.4 & 782.6 & 832.9 & 847.3 & 799.0 \\
\quad + Merge & 808.2 & 810.5 & 816.6 & 866.9 & 881.4 & 833.1 \\
\bottomrule
\end{tabular}

\label{tab:validation-results}
\end{table}
\vspace{-3em}

\section{Machine configuration}
\label{sec:machine}
All experiments were conducted on a machine running Windows~11 Pro
equipped with an AMD Ryzen~7 PRO 5750G processor (8 cores / 16 threads, 3.8~GHz base, 4.6~GHz boost, Zen~3 architecture)
and 128~GB of DDR4 RAM.
The JVM heap was capped at 100~GB (\texttt{-Xmx100g}).

\end{document}